# Two-Dimensional Lattice-Gas Model for Methane Clathrate Hydrates: Comparative Analysis with Experiments and Three-Dimensional Simulations

Julián Juan[a], María E. Pronsato[b], Antonio J. Ramirez-Pastor[a] and Pablo Longone [a,1*]

[a]Departamento de Física, Instituto de Física Aplicada (INFAP), Universidad Nacional de San Luis, CONICET, Ejército de los Andes 950, D5700HHW San Luis, Argentina.

[b] Instituto de Física del Sur (IFISUR), Departamento de Física, Universidad Nacional del Sur (UNS), CONICET, Av. L. N. Alem 1253, B8000CPB-Bahía Blanca, Argentina

**Abstract**

Methane clathrate hydrates, particularly those with an sI structure, are significant due to their potential as energy resources and their impact on gas pipelines. In this study, a two-dimensional (2D) lattice-gas model is employed to investigate the main thermodynamic properties of methane clathrate hydrates. The proposed framework is validated through comparison with experimental data and more advanced three-dimensional (3D) simulations. Adsorption isotherms, dissociation enthalpy and phase stability of the sI structure are evaluated using Monte Carlo (MC) simulations in the grand canonical ensemble. The 2D adsorption isotherms closely align with both experimental data and 3D simulations, demonstrating the 2D model's ability to precisely represent both rigid and flexible sI structures. The dissociation enthalpy calculated using our approach (76.4 kJ/mol) excellently matches the experimental value (78 kJ/mol), confirming the model's accuracy. Furthermore, the phase diagram obtained from the Clausius-Clapeyron equation shows very good agreement with experimental data between 260 and 290 K, though deviations are observed above 290 K. These findings underscore the effectiveness and robustness of the 2D model in studying methane clathrate hydrates and suggest its potential applicability for investigating other guest species and hydrate structures.

---

1* Corresponding author: plongone@unsl.edu.ar

## 1. Introduction

Clathrate hydrates are inclusion compounds consisting of a three-dimensional (3D) crystalline lattice of frozen water cavities that host molecules of gas with low dipole moments [1]. Typically, gas clathrate hydrates form under extreme conditions such as high pressures and low temperatures. These regular lattices can form different structural types depending on the nature of the guest species they host. Gas molecules like methane, ethane, and carbon dioxide form type I hydrates, whereas larger molecules preferentially form type II (propane, isobutane) and sH (cyclohexane, cycloheptane) structures [1,2]. In the particular case of methane clathrate hydrate with sI structure, each unit cell contains two dodecahedral cavities ($5^{12}$) (small cavity) and six hexakaidecahedron cavities ($5^{12}6^2$) (large cavity). The symbols in parentheses correspond to the Jeffrey notation of cavities [3]. Methane hydrates are crucial in various scientific fields related to planetology and are also considered one of the primary natural resources for the energy industry [4,5,6]. In recent years, large methane deposits have been found in natural geological environments such as ocean depths and Arctic permafrost regions. It has also been estimated that there are more energy resources in these hydrates than in fossil fuels [7,8]. Furthermore, the industry and scientific community have paid significant attention to $CH_4$ hydrates due to the importance of gas pipeline flow obstruction problems caused by the formation of large masses of methane clathrate hydrates [9], as well as $CH_4$ production during hydrate dissociation in geological formations [10-18].

Currently, there is a limited amount of experimental studies on storage capacity (occupancy), stability, and formation conditions for hydrates such as methane. Consequently, theoretical models capable of predicting the thermodynamic behavior of hydrates have been developed. One of the most important theoretical models for predicting the formation and occupancy conditions of clathrate hydrates is the van der Waals and Platteeuw (vdWP) theory [19]. This theory arises from the combination of statistical mechanics and classical adsorption theory (Langmuir model). Among the most important characteristics of the vdWP approach are: (i) the interactions between guest molecules and the water lattice are relatively weak and limited to the nearest neighbors (NN) of the gas molecules; (ii) the interactions between guest molecules in adjacent cavities are neglected; and (iii) only simple occupancy of guest molecules per cavity is considered. Consequently, the behavior of each guest molecule is independent of the presence of other guest molecules. Over the years, various modifications have been made to the vdWP model, including factors such as multiple occupancy and cavity flexibility [20-25]. However, the vdWP theory in combination with different equations of state is the most widely used theory by the scientific and academic community to predict formation and occupancy conditions in gas clathrate hydrates [26].

In recent years, numerous studies have employed computational simulations, including Molecular Dynamics (MD) and Monte Carlo (MC) techniques, to predict the storage capacity, stability, and formation conditions of clathrate hydrates for various species and structures [27-43]. In the case of methane hydrates (which is the topic of this paper), various studies have been conducted using MC simulations [33-46]. Our research group has developed a 2D lattice-gas model along with MC simulations to investigate quantities such as cavity occupancy, the degree of sI structure distortion, and their relationship with Helmholtz free energy [44-46]. The studies in Refs. [44, 45] demonstrated the existence of minimum cavity deformation values, which correspond to minimum free energy values. For example, in the case of $CH_4$, this occurs when there is one molecule per cavity (this is $\theta_{cavity} \approx 1.0$). Additionally, qualitative phase diagrams were obtained for various guest species, including ($CO_2$, $CH_4$ and $C_2H_4$). Later, Ref. [46] revealed that the $CO_2$-$CH_4$ exchange process within the sI structure is driven not only by energy magnitudes but also by an entropic behavior, characterized by a spontaneous increase in $CO_2$ entropy compared to $CH_4$. Although these lattice-gas models are discrete, simplified, and operate in two spatial dimensions, they offer several advantages. Firstly, they require very few parameters. Secondly, their computational cost is relatively low. Thirdly, they allow for the modeling of sI structure distortion (or deformation) through processes involving the diffusion of monomers previously adsorbed within the lattice. Lastly, the triangular geometry of the lattice, which serves as the foundational base of the sI structure, preserves both the symmetry of the sI hydrate structure and the relative sizes of guest molecules. As a result, these models seem to effectively capture the essence of real clathrate hydrate systems. It is worth noting that lattice-gas models have demonstrated significant potential for describing a wide range of physical and other field phenomena [47]. In the early twentieth century, two important research methods related to the lattice-gas model were developed: the Bragg-Williams approximation [48] and the quasi-chemical approximation [48]. Since then, there have been significant advancements in developing accurate and powerful approximations, such as the real-space renormalization group technique [49], the transfer matrix method [50] and finite-size scaling theory [51]. As a result of these developments, lattice-gas theory, originally proposed to model ferromagnetism [52-54], has become one of the most active research areas for studying phase transitions and critical phenomena [47].

In this work, building on our previous papers [44-46], we take a further step in validating our 2D lattice-gas model by conducting an extensive comparison with experimental results and advanced 3D simulations focused on the problem of methane clathrate hydrate. To achieve this, the article is divided into two stages. The first is devoted to the comparative study of storage capacity through adsorption isotherms. This initial stage is supplemented with simple Density Functional Theory (DFT) calculations [55, 56], aimed at reliably determining the lattice parameters in the sI structure. The second stage involves determining the dissociation enthalpy and constructing the pressure-temperature phase diagram. This study enables the identification of stability and instability regions, which are then compared with experimental data from the literature.

The present paper is a natural extension of our previous research in the area of clathrate hydrates. The scheme developed here could be applied and extended to other structures and different guest species such as $CO_2$, $C_2H_4$, and $N_2$. This document is organized as follows: The details of the model and MC simulation scheme are given in Section 2. The results are presented and discussed in Section 3. Finally, the main conclusions are mentioned in Section 4.

## 2. Models and methods

2.1. Lattice-gas model: the sI hydrate structure and $CH_4$ as a guest species.

In this section, we present a simplified overview of the 2D lattice-gas model for the sI structure of clathrate hydrates and the guest molecule $CH_4$. This model, which has been developed and explained in detail in our previous works [44-46], is analogous to well-known adsorption-desorption models with multiple-site occupancy [57-63].

Let us consider a portion of the 3D sI structure, formed by 8 unit cells (2 x 2 x 2), where each unit cell contains two dodecahedral cavities ($5^{12}$) (small cavity) and six tetrakaidecahedral cavities ($5^{12}6^2$) (large cavity). Then, by cutting through planes that pass through its center, the sI structure is divided into two symmetrical parts [64, 65].

Each part has faces equivalent to those of the cube formed by the eight unit cells (2 × 2 × 2). Due to this symmetry, the idea of modeling the sI structure as a 2D lattice-gas model emerges. The procedure involves discretizing one of the many equivalent faces of the sI structure and then mapping it onto a 2D triangular lattice. In Fig. 1(a), one of the equivalent sI faces can be observed, and water molecules can be identified on a plane forming chains of hexagons (continuous blue lines). The spaces between the hexagon chains correspond to the small and large cavities of the 2D sI structure (dashed lines). Meanwhile, in Fig. 1(b), the mapping of the actual sI face depicted in Fig. 1(a) (enclosed with a white dashed rectangle) to a regular triangular lattice of sites can be observed. This lattice, with a high degree of connectivity ($c=6$), presents occupied and unoccupied sites. The sites occupied by monomers (objects occupying a single site, $k=1$, blue spheres) represent the water molecules forming the structure of the parallel chains of hexagons along the length $L_1$ and width $L_2$ of the lattice. The $O$ sites between hexagons do not directly correspond to individual water molecules; instead, they represent vacant sites that could potentially be occupied by water molecules displaced from their equilibrium positions, thus deforming the sI structure. The empty sites between the chains correspond to the small (three sites) and large (five sites) cavities. The resulting 2D structure contains one small cavity for every two large cavities, see Fig. 1(b). The ratio of small to large cavities for real sI hydrates is one small cavity for every three large ones. These differences represent a limitation and must be taken into account when comparing our model with experiments, or even with other theoretical models considering the real structure of the hydrate. However, as already shown in our previous research [44-46], interesting information can be extracted from the simple 2D lattice-gas approximation.

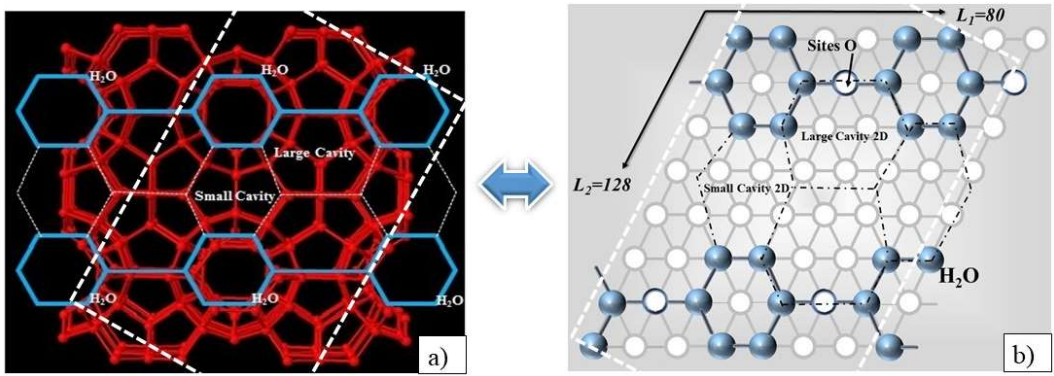

**Fig. 1:** (a) The surface face of the sI structure depicts water molecules at the ends of the hexagons, while empty spaces delineated by white lines represent small and large cavities. (b) The 2D

triangular lattice-gas model comprises $L_1 \times L_2$ sites, featuring chains of previously adsorbed monomers. The spaces between these chains give rise to both large and small cavities. A slanted rectangle outlined with white dashed line frames a portion of the surface, which is mapped onto part (b).

We now move on to the description of the adsorbate. The $CH_4$ molecule possesses a tetrahedral geometry [66], where each face of the tetrahedron is an equilateral triangle. Whenever the tetrahedron situates or adsorbs onto a surface, it does so by settling its triangular base as the only possible form, see Fig. 2(a). Due to this geometric property, we represent the 2D methane molecule using a base with three vertices forming an equilateral triangle. The length of each side of the triangle is chosen to be equal to the lattice constant $a$. Accordingly, each $CH_4$ guest molecule occupies three sites when adsorbed on the lattice ($k=3$). In Fig. 2(b), the guest species ($CH_4$) can be observed adsorbed on the triangular lattice forming the 2D sI structure. In the case of the figure, the sI structure exhibits a distorted state configuration (water molecules are displaced from their original positions).

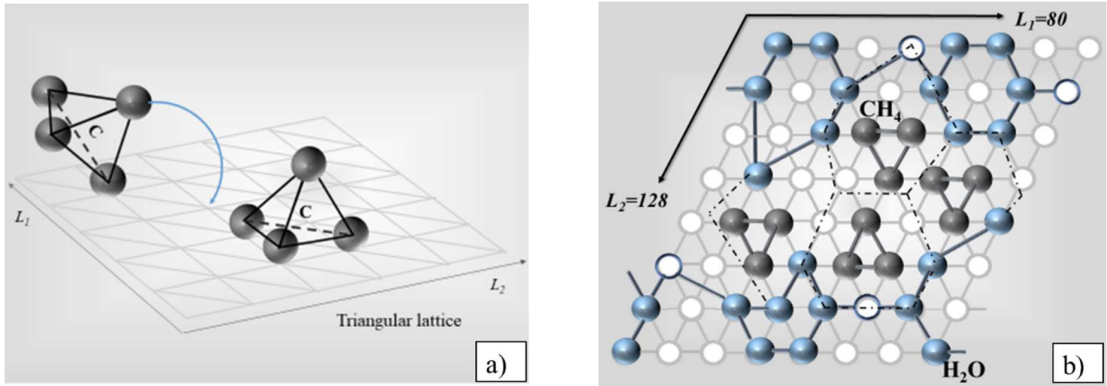

**Fig. 2:** (a) Adsorption sequence for a $CH_4$ molecule forming an equilateral triangle on a triangular lattice. (b) Typical configuration state of the guest species $CH_4$ and water molecules. Note the distortion of the sI structure. The water molecules ($k=1$) form the parallel chains of hexagons. Some of them remain in their original position while others have shifted to neighboring sites. The $CH_4$ molecules ($k=3$) occupy both small and large cavities.

To complete our analysis, it is crucial to determine the lateral interactions among the system components. For this purpose, we adopt the Lorentz-Berthelot combination rule [23, 24, 67, 68], widely employed in this field of study,

$$\varepsilon_{ij} = \sqrt{\varepsilon_{ii}\varepsilon_{jj}}. \tag{1}$$

Here, $\varepsilon_{ij}$ represents the interaction between two molecules, $i$ and $j$. Eq. (1) enables the prediction of mixture properties ($\varepsilon_{ij}$) based on data from pure components ($\varepsilon_{ii}$ and $\varepsilon_{jj}$).

In our lattice-gas model, we aim to determine the lateral interaction energy $w_{ij}$ between two NN monomers from different molecules (whether of the same or different species). This can be readily calculated using Eq. (1) by dividing $\varepsilon_{ii}$ and $\varepsilon_{jj}$ by the number of monomers (or units) comprising molecules $i$ and $j$, respectively. Thus

$$w_{ij} = \sqrt{\frac{\varepsilon_{ii}}{k_i}\frac{\varepsilon_{jj}}{k_j}}, \tag{2}$$

where $k_i$ ($k_j$) denotes the number of units or monomers that form the molecule $i$ ($j$). In this case, $i$ and $j$ are elements of the set {$CH_4$, $H_2O$}, with $k_{CH4} = 3$ and $k_{H2O} = 1$ [see Fig. 2(b)]. Additionally, the Lennard-Jones interaction parameters for the pure species are as follows: $\varepsilon_{CH4\text{-}CH4}/k_B = 148.1$ K, and $\varepsilon_{H2O\text{-}H2O}/k_B = 102.1$ K, where $k_B$ represents the Boltzmann constant (see Table 2 in Ref. [69]).

In some instances, Eq. (1) has been modified by introducing the binary interaction parameter $l_{ij}$, allowing Eq. (2) to be rewritten as

$$w_{ij} = l_{ij}\sqrt{\frac{\varepsilon_{ii}}{k_i}\frac{\varepsilon_{jj}}{k_j}}. \tag{3}$$

The value of $l_{ij}$ can be correlated with experimental cavity occupancy data. As in Refs. [24, 25, 44-46], $l_{ij}$ is commonly assumed to be unity ($l_{ij} = 1$). Furthermore, for the $CH_4$-$H_2O$ interaction, $l_{ij}$ falls within a range close to unity ($1 \leq l_{ij} \leq 1.07$) [70]. The authors in Ref. [70], using the TIP4P/2005 model and MC simulations in *NVT*, determined that $l_{ij} = 1.07$ provides better agreement with experiments, describing the excess chemical potential of methane hydrate clathrates sI in terms of "effective" polarizability. However, in our lattice gas model or coarse-grained simplification, there is no significant difference in using a value of $l_{ij}$ between 1 and 1.07 when introducing lateral interactions.

Finally, the implementation of the precursor model described in this section involves the omission of some experimental features present in real clathrates, primarily the reduction of one degree of freedom. However, even when simplifying the problem to a discrete 2D lattice, the proposed model preserves the inherent symmetry of the sI hydrate structure and the relative sizes of the guest molecules. These characteristics seem to capture the essence of real systems. In section 3, a quantitative comparison with 3D simulations, theoretical models, and literature experiments will be presented, paving the way for a new discussion on the potential demonstrated by our 2D model.

2.2. MC simulation scheme and thermodynamic quantities

As discussed in Section 2.1, we have represented the equivalent faces or planes of the sI structure using a 2D lattice-gas model with triangular geometry. Due to the topography and connectivity of the triangular lattice, the pre-adsorbed monomers (water molecules) form chains of hexagons. The empty sites between these hexagon chains give rise to the formation of both large and small cavities. Additionally, the adsorbed $CH_4$ molecule is modeled by occupying three sites, thereby forming a set of adjacent monomers with the following sequence: After randomly placing the first monomer at the initial adsorption site, three possible directions for the occupation of the second monomer to its NNs are drawn. Once the second neighbor is occupied, the third monomer occupies one of the two NNs to the first monomer [see Fig. 2(b)].

The surface layer of the hydrate is represented by an $M = L_1 \times L_2$ matrix of adsorption sites with periodic boundary conditions. However, the number of active adsorption sites is $M^* = L_1 \times L_2 - (N_{H2O} + N_O)$, where $N_{H2O}$ ($N_O$) is the number of $H_2O$ molecules ($O$ sites) in the sI structure. Therefore, our system consists of two components. First, there are $N_{H2O}$ water molecules that can be considered as pre-adsorbed on the lattice. These water molecules can only move to neighboring sites (including $O$ sites), so their total number remains constant throughout the process. Second, there are $N$ methane molecules adsorbed on $M^*$ sites at a given temperature $T$ and chemical potential $\mu$. The $CH_4$ guest molecules adsorb and desorb as independent units, without accounting for possible dissociation. It is useful to express the system's energy, $E$, in terms of these two components:

$$E = E^G + E^W, \tag{4}$$

where $E^G$ is the energy of the adsorbed guest phase and $E^W$ is the energy associated to the aqueous lattice.

As usual in lattice-gas models, an occupancy variable $c_i$ is introduced. In our case, this variable can take four distinct values: $c_i = 0$ if site $i$ is vacant, $c_i = 1$ if site $i$ is occupied by a water molecule, $c_i = 2$ if site $i$ is occupied by a unit belonging to a $CH_4$ molecule, and $c_i = 3$ if site $i$ is an $O$ site. Under these considerations, $E^G$ and $E^W$ can be written as

$$E^W = \sum_{(i,j)} [w_{11} \delta_{c_i,1} \delta_{c_j,1} + w_{12}(\delta_{c_i,1} \delta_{c_j,2} + \delta_{c_i,2} \delta_{c_j,1})], \qquad (5)$$

and

$$E^G = \sum_{(i,j)} [w_{22} \delta_{c_i,2} \delta_{c_j,2} + w_{12}(\delta_{c_i,1} \delta_{c_j,2} + \delta_{c_i,2} \delta_{c_j,1})] - 3w_{22} N, \qquad (6)$$

where $(i,j)$ represents pairs of NN sites and $\delta$ is the Kronecker delta function; $w_{11}$ is obtained from Eq. (3) and represents the *NN* interaction between water molecules ($H_2O$-$H_2O$); $w_{22}$ [Eq. (3)] is the *NN* interaction between guest molecules ($CH_4$-$CH_4$) and $w_{12}$ [Eq. (3)] corresponds to the *NN* interaction between a water molecule and a guest molecule ($H_2O$-$CH_4$). $w_{11}$, $w_{12}$ and $w_{22}$ are considered attractive interactions ($w_{11}$, $w_{12}$, $w_{22}$ <0). The term $3w_{22}N$ is subtracted in Eq. (6) since the summation over all the pairs of *NN* sites overestimate the total energy by including $3N$ internal bonds belonging to the $N$ adsorbed guest molecules. Finally, it can be observed that the $O$ sites ($c_i = 3$) were not taken into account in Eqs. (4-6) because they do not energetically interact in the system (see Refs. [43-45]).

The adsorption-desorption process of $CH_4$ on the 2D sI structure is explored via MC simulations in the grand canonical ensemble ($\mu VT$ ensemble)[2], employing an adsorption-desorption algorithm with multisite occupancy [57-63]. This algorithm offers three fundamental pathways for altering the state of the adsorbed guest phase: adsorbing a guest molecule onto the surface ($\Delta N = +1$), and desorbing a guest molecule from the adsorbed phase ($\Delta N = -1$). The corresponding transition probabilities are determined using the

---

[2] In our simplified model, the volume $V$ corresponds to the number of available sites for adsorption $M^*$, a standard assumption in 2D lattice gas models.

Metropolis rule [71], expressed as $W = \min\{1, \exp[-\beta(\Delta E^G - \mu \Delta N)]\}$, where $\beta = 1/(k_B T)$ and $\Delta E^G = E_f^G - E_i^G$ represents the difference between the energies of the final and initial states of the adsorbed guest phase [refer to Eq. (6)]. Each elementary adsorption/desorption step is accompanied by an attempt to move a water molecule to one of its neighboring positions. This displacement attempt is accepted or rejected based on the Metropolis scheme as well [71]. In this case the corresponding transition probability can be written as $W = \min\{1, \exp[-\beta(\Delta E^W)]\}$, where $\Delta E^W = E_f^W - E_i^W$ represents the difference between the energies of the final and initial states of the aqueous lattice [refer to Eq. (5)]. As mentioned above, the quantity of water molecules remains constant throughout the simulation ($\Delta N_{H2O} = 0$).

One MC step (MCS) corresponds to a set of $M$ consecutive adsorption-desorption-displacement attempts. The equilibrium state can be well reproduced after discarding the first $m_0 = 10^7$ MCSs. Then, averages have been evaluated over $m_1 = 10^7$ successive configurations. In our simulations, we varied the chemical potential and monitored the surface coverage $\theta$, and the cavity coverage $\theta_{cavity}$. These quantities are obtained as simple averages,

$$\theta = \frac{3\langle N_{CH4} \rangle}{M^*}, \qquad (7)$$

and

$$\theta_{cavity} = \frac{\langle N_{CH4} \rangle}{n_{cavity}}, \qquad (8)$$

where $n_{cavity}$ is the number of cavities (small and large) in the lattice, and the bracket $\langle ... \rangle$ denotes the average over $m_1$ MC simulation runs after equilibrium is settled.

The free energy of the adsorbed guest phase $F$ is calculated by using the thermodynamic integration method [72]. The method in the grand canonical ensemble relies on the integration of the chemical potential $\mu(\theta)$ coverage along a reversible path between an arbitrary reference state and the desired state of the system. Thus, for $N$ guest molecules on $M^*$ lattice sites,

$$F(N, M^*, T) = F(N_0, M^*, T) + \int_{N_0}^{N} \mu dN'. \tag{9}$$

In our case, the determination of the free energy in the reference state, $F(N_0, M^*, T)$, is trivial [$F(N_0, M^*, T) = 0$ for $N_0 = 0$]. Note that the reference state, $N \to 0$, is obtained for $\mu / k_B T \to -\infty$. Finally, Eq. (9) can be written in terms of intensive variables,

$$\beta f = \int_0^\theta \frac{\mu(\theta')}{k} d\theta'. \tag{10}$$

The dissociation enthalpy $\Delta H_{diss}$ was also calculated. The methodology used is based on the proposal by E. D. Sloan [73]. This is a preliminary approximation for simple species, starting from the heat required for the dissociation of the clathrate hydrate into its individual components according to the following equation; *water + gas ↔ clathrate hydrate*. Thus, the quantity $\Delta H_{diss}$ depends on two energetic terms

$$\Delta H_{diss} = \Delta H_{st}^G + \Delta H_{st}^W, \tag{11}$$

where $\Delta H_{st}^G$ is isosteric enthalpy of adsorption for the guest species (or adsorbate) and $\Delta H_{st}^W$ represents the contribution of the water lattice. In our simulations, the adsorbed molecules remain fixed in their equilibrium positions. Accordingly, the translational, rotational and vibrational degrees of freedom of the molecules in the adsorbed phase are ignored in Eq. (11).

The isosteric enthalpy $\Delta H_{st}^G$ is calculated following the standard procedure in grand canonical MC simulations [74],

$$\Delta H_{st}^G = -\left[\frac{\partial \langle E^G \rangle}{\partial \langle N \rangle}\right]_T = -\frac{\langle E^G N \rangle - \langle E^G \rangle \langle N \rangle}{\langle N^2 \rangle - \langle N \rangle^2}. \tag{12}$$

In the last equation, the derivative of the energy of the adsorbed guest phase with respect to the number of guest molecules is expressed as fluctuations in the grand canonical ensemble. As in previous Eqs. (7) and (8), $\langle ... \rangle$ means the average over the $m_1$ MC simulation runs.

Regarding $\Delta H_{st}^{W}$, this quantity represents the contribution to the total enthalpy resulting from the dynamics of the hexagon chains, which become deformed during the adsorption process of the guest molecules. To calculate $\Delta H_{st}^{W}$, it is necessary to consider the energy changes in the water lattice structure $\Delta E_{i}^{W}$ involved in successful attempts to move a water molecule to a neighboring site. The total number of these successful attempts, $N_{sda}$, must also be considered. The MC procedure is as follows. After reaching equilibrium, following the discarding of the first $m_0 = 10^7$ MCSs, the enthalpy $\Delta H_{st}^{W}$ is obtained by averaging the change in energy $\Delta E_{i}^{W}$ in successful attempts of diffusion or displacement of water molecules during the adsorption process,

$$\Delta H_{st}^{G} = -\left(\frac{\sum_{i=1}^{N_{sda}} \Delta E_{i}^{W}}{N_{sda}}\right). \tag{13}$$

By combining Eqs. (11-13), the dissociation enthalpy $\Delta H_{diss}$ is obtained. Then, we apply the Clausius-Clapeyron equation (Pressure $P$ vs. Temperature $T$):

$$\frac{dP}{d\left(\frac{1}{T}\right)_{\theta \approx 1}} = -\frac{\Delta H_{diss}}{ZR}, \tag{14}$$

where $Z$ is the compressibility factor and $R$ is the universal gas constant. The gas phase is considered ideal; hence $Z$ is set to unity. This allows us to calculate the first-order transition points, which experimentally correspond to the three-phase coexistence line (Liquid$_{H2O}$-Hydrate-Gas or Ice$_{H2O}$-Hydrate-Gas) within the temperature range of 257-294 K.

3. **Results and discussion**

In this section, we present the results for CH$_4$ adsorption using the 2D lattice-gas model with periodic boundary conditions and the MC simulation scheme described in Sections 2.1 and 2.2. The lattice size used is $L_1 \times L_2 = 80 \times 128$ (3072 H$_2$O molecules and 512 O sites), which accurately represents the sI structure and is large enough to avoid finite-size effects. The chemical potential $\beta\mu$ (in $\beta$ units) was used as a control parameter. In this work, the results are divided into two parts. First, our 2D adsorption isotherms are compared with experiments and more advanced 3D simulations from the literature (Subsection 3.1). In the second part (Subsection 3.2), the calculations of dissociation enthalpy are shown, and consequently, the calculation of the $P$ vs $T$ phase diagram is performed to again make a quantitative comparison with the experimental and theoretical data from the literature.

### 3.1. *Adsorption isotherms*

In Section 2.2 and Eq. (8), the cavity coverage $\theta_{cavity}$ was defined for a fixed value of chemical potential $\beta\mu$ and temperature $T$. Varying the chemical potential, the adsorption isotherm curve for the CH$_4$ clathrate hydrate on the sI structure in 2D ($\theta_{cavity}$ vs $\beta\mu$) can be calculated.

Before starting the comparative analysis, and to facilitate comparison with experiments and 3D simulations from the literature, it is convenient to write the theoretical isotherms in terms of the pressure. To do this, we introduce the equation that relates the chemical potential to the pressure for a lattice gas [69],

$$\mu = k_B T \ln(P) - k_B T \ln\left(\frac{k_B T}{\Lambda^3}\right) + k_B T \ln\left(\frac{a}{\Lambda}\right)^b \qquad (15)$$

where $a$ is the lattice constant, $b$ is related to the degrees of freedom of the guest molecule (for example, $b = 3$ in 3D space), $P$ is the pressure, and $\Lambda$ is the thermal de Broglie wavelength.

In this study, the lattice constant $a$ was calculated through DFT calculations using a sI clathrate hydrate in three dimensions for the CH$_4$ species at 270K and 100% cavity occupancy. These calculations were performed using the Vienna Ab-initio Simulation Package (VASP) [75-77]. The Perdew-Burke-Ernzerhof (PBE) functional in the generalized gradient approximation (GGA-PBE) was used for the exchange-correlation functional, with an energy cut-off of 520 eV and a single k-point for the Brillouin zone integrations because of the large size of the cell, considering a Γ-centered mesh [78]. Convergence criteria 1x10$^{-4}$ eV and 2x10$^{-2}$ eV/Å was used for energy and forces, respectively. Van der Waals correction were taken into account with the DFT-D3 Grimme method [79]. Ab initio Molecular Dynamics calculations were performed in an NVT ensemble [80] with a 1 fs timestep and 2 ps of total time.

The obtained value was $a = 2.57 \times 10^{-10}$ m. On the other hand, $b$ was taken as an adjustable parameter.

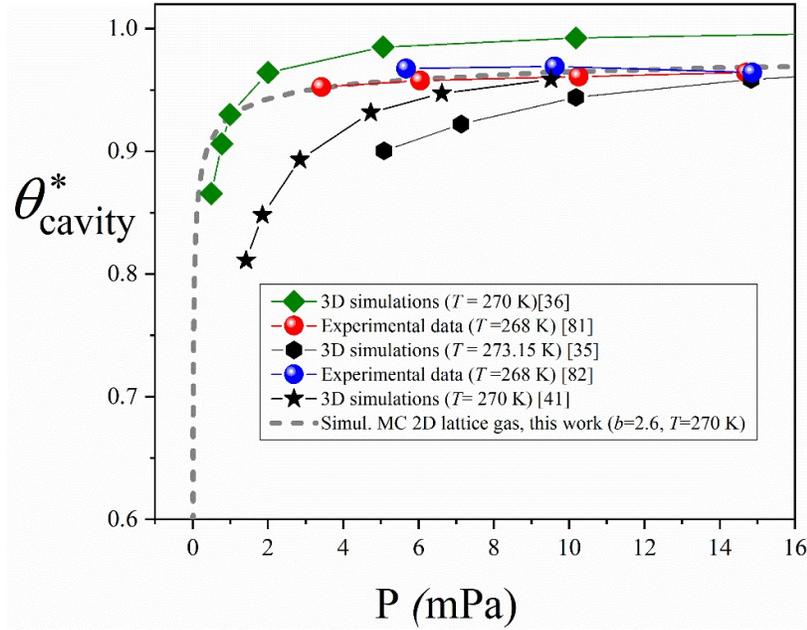

**Fig. 3:** Comparison between our 2D adsorption isotherm (gray dashed line), experimental adsorption isotherms in Refs. [81, 82] (red and blue spheres, respectively), and 3D simulations from Ref. [35] (black hexagons), Ref. [36] (green diamonds) and Ref. [41] (black stars). The parameter values used in the curves shown in the figure are as follows: 2D model ($b = 2.6$ and $T = 270$ K); Refs. [81, 82] ($T = 268$ K); Ref. [35] (T = 273.15 K) and Refs. [36, 41] (T = 270 K).

Figure 3 shows the adsorption isotherm obtained using our 2D model (dashed line) in comparison with experiments and more advanced 3D and simulations [35, 36, 41, 81, 82]. The 2D adsorption isotherm is presented in terms of the normalized cavity coverage $\theta^*_{cavity}$ ($0 \leq \theta^*_{cavity} \leq 1$) and pressure $P$ (mPa) [Eq. (15) with $a=2.57 \times 10^{-10}$ m]. The temperature was set to $T = 270$ K and the range of $\theta^*_{cavity}$ shown in the figure varies from 0.6 to 1.0. Figure 3 also includes (a) experimental data from Ref. [81] (red spheres) and Ref. [82] (blue spheres). The occupancy experiments, conducted by Qin et al., were obtained from Rietveld refinement using high-resolution X-ray diffraction patterns and Raman spectroscopy measurements at temperature $T = 268$ K; and (b) 3D MC simulations using NCSR at 273.15 K from Ref. [35] (black hexagons) and 3D MC simulations in the grand canonical ensemble at 270 K from Ref. [36] (green diamonds) and Ref. [41] (black stars).

It is important to note that although the simulations and experiments are not conducted at exactly the same $T$, the comparison is valid because in a small temperature interval of 268-273 K, cavity occupancy is not expected to change significantly. For this reason, the working $T$ used in the 2D simulations to achieve a comparison with the experiments is 270 K.

An excellent agreement is observed between the experimental data [81, 82] and the 2D model predictions for $b = 2.6$. The value of the parameter $b$ obtained from the fit (slightly less than 3) suggests the presence of confinement in the system. In fact, as shown in Figs. 1(b) and 2(b), the water molecules and the $O$ sites form parallel chains of hexagons, confining the system along the $L_2$ direction. In other words, the available sites for $CH_4$ adsorption form channels with finite length along the $L_2$ direction. This behavior was already observed in our previous work for the case of ideal 2D clathrate hydrates [63].

The differences between 2D simulations and experimental results can be much easily rationalized with the help of the percentage reduced coverage $AD\ \%$, which is defined as,

$$AD\ \% = \left| \frac{\theta_{cavity}^{exp} - \theta_{cavity}^{*}}{\theta_{cavity}^{exp}} \right| \times 100, \quad (16)$$

where $\theta_{cavity}^{exp}$ represents the cavity coverage obtained from experiments. Each pair of values ($\theta_{cavity}^{*}$, $\theta_{cavity}^{exp}$) is obtained at fixed $P$. The $AD\ \%$ values resulting from the comparison of our 2D adsorption isotherm with the two experiments conducted by Qin and collaborators [81, 82] are summarized in Table 1.

Table 1: Performance of the 2D lattice-gas model presented in Section 2.1 for methane clathrate hydrate compared with experiments from the literature.

| $\theta^*_{cavity}$ (this work) | $\theta^{exp}_{cavity}$ [81] | $P/mPa$ | $T/K$ | $AD$ % |
|---|---|---|---|---|
| 0.95258 | 0.9535 | 3.44 | 270 | 0.09649 |
| 0.9536 | 0.958 | 6.01 | 270 | 0.01044 |
| 0.9579 | 0.9607 | 10.27 | 270 | 0.30186 |
| 0.95807 | 0.964 | 14.7 | 270 | 0.3112 |
| $\theta^*_{cavity}$ (this work) | $\theta^{exp}_{cavity}$ [82] | $P/mPa$ | $T/K$ | $AD$ % |
| 0.968 | 0.9565 | 5.648 | 270 | 1.18802 |
| 0.9698 | 0.963 | 9.588 | 270 | 0.70118 |
| 0.9643 | 0.96709 | 14.87 | 270 | 0.28933 |

Table 1 quantitatively reflects the excellent agreement with the experiments from Refs. [81, 82]. For the first experiment [81], the *AD* % value for each isotherm point is very low, reaching a maximum value around 0.31 % in the high-pressure range (*P* = 10-14 mPa). However, for the second experiment [82], there is a slightly higher error compared to the first experiment, such as 1% (1.18802) for a lower *P* range (5.6-9.6 mPa), then the error decreases with increasing *P*. The discrepancies between the experiments and our simulations are related to the uncertainty of the non-stoichiometry of methane clathrate hydrates [1], as well as to the fact that the exact formation conditions are not fully known. While we recognize the challenging and varied contexts in which these experiments were conducted, we believe these studies offer excellent, real, and direct evidence of high cavity coverage for methane at different pressure values.

To conclude with the analysis of Fig. 3, it can be observed that the 3D simulations [35, 36, 41] present a reasonably close agreement to the experimental curves in the high-pressure range of 10 to 14.5 mPa. However, for low *P*, these simulations deviate significantly from the experimental isotherms.

We continue our study by comparing the 2D model with 3D MC simulations by Sizov and Piotrovskaya [34]. The calculations in Ref. [34] were performed using the SPC/E model for water [83] and the UA model for methane molecules [84], for both flexible and rigid sI structures at $T = 200$ K. In the case of the 2D model, the procedure employed is similar to that used in Fig. 3. Specifically, once the cavity coverage [Eq. (8)] is determined as a function of the chemical potential, the corresponding $P$ values are calculated using Eq. (15), with $b$ as the adjustable parameter. In this case, the resulting values of b were $b = 2.70$ for the rigid structure and $b = 2.81$ for the flexible structure. Additionally, the simulations for the rigid structure do not account for the mobility of water molecules (the aqueous lattice remains fixed during the simulations). The results of the comparison are shown in Fig. 4. 2D and 3D adsorption isotherms are depicted in dashed lines and open circles, respectively. Red and blue symbols represent data for rigid and flexible structure, respectively. All curves in the figure have been obtained at $T = 200$ K.

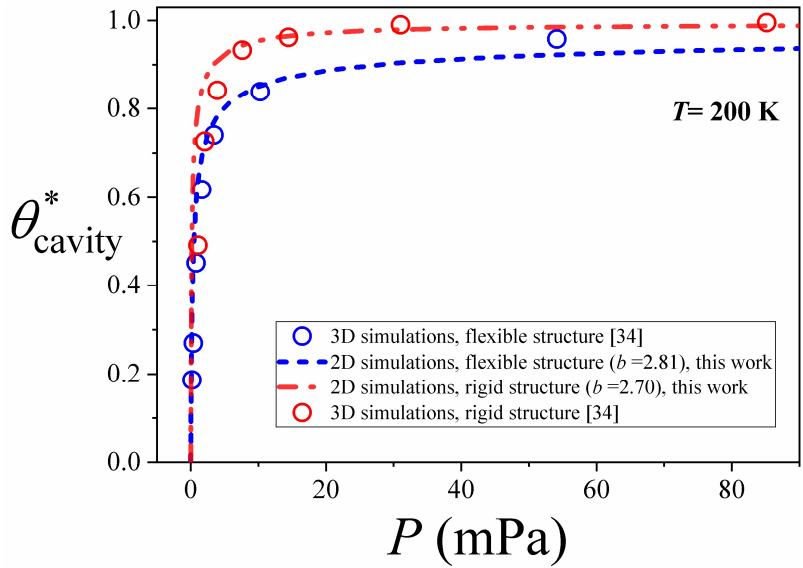

**Fig. 4:** Comparison between 2D adsorption isotherms (dashed lines) and those obtained in Ref. [34] using 3D MC simulations (open circles). Red and blue symbols represent data for rigid and flexible structure, respectively. All curves in the figure have been obtained at 200 K. In the case of the 2D model, the parameter $b$ was set to 2.70 for the rigid lattice and 2.81 for the flexible lattice.

As shown in Fig. 4, there is very good agreement between our 2D model and the 3D simulations by Sizov and Piotrovskaya [34] for both flexible and rigid sI structures across the entire P range (0–90 mPa). It is interesting to examine the behavior of parameter b, which is a measure of the dimensionality of the clathrate structure. The $b$ values obtained from the fits in Fig. 4 are less than 3, indicating the presence of confinement effects [63]. These effects are more pronounced in the rigid structure ($b = 2.7$) than in the flexible structure ($b = 2.81$). This behavior can be attributed to the fact that in a rigid structure, the elements forming the chains of hexagons remain fixed, restricting the degrees of freedom of the methane molecules.

The differences between the 2D and 3D simulations can be quantified using the previously defined percentage reduced coverage $AD$ % [Eq. (16)]. In this case, $\theta_{cavity}^{exp}$ (experimental points) is replaced by $\theta_{cavity}^{3Dsim}$ (3D simulation points). The resulting $AD$ % values are collected in Table 2.

Table 2: Performance of the 2D lattice-gas model presented in Section 2.1 for sI methane clathrate hydrate compared to 3D MC simulations at 200 K (rigid structure and flexible structure) [34].

| Rigid structure | | | | |
|---|---|---|---|---|
| $\theta_{cavity}^{*}$ (this work) | $\theta_{cavity}^{3Dsim}$ [34] | $P$ / mPa | $T$ / K | $AD$ % |
| 0.1865 | 0.18648 | 0.21 | 200 | 0.0089 |
| 0.2697 | 0.26969 | 0.40 | 200 | 0.00319 |
| 0.4544 | 0.45084 | 0.80 | 200 | 0.79035 |
| 0.61747 | 0.61751 | 1.63 | 200 | 0.00668 |
| 0.77 | 0.74051 | 3.46 | 200 | 3.98178 |
| 0.85149 | 0.83845 | 10.30 | 200 | 1.55504 |
| 0.923 | 0.9581 | 54.17 | 200 | 3.66305 |
| Flexible structure | | | | |
| $\theta_{cavity}^{*}$ (this work) | $\theta_{cavity}^{3Dsim}$ [34] | $P$ / mPa | $T$ / K | $AD$ % |
| 0.4911 | 0.49125 | 1.06 | 200 | 0.03053 |
| 0.7235 | 0.72651 | 2.07 | 200 | 0.41445 |
| 0.898 | 0.891 | 3.92 | 200 | 0.781 |
| 0.945 | 0.93287 | 7.63 | 200 | 1.29982 |
| 0.9651 | 0.96177 | 14.5 | 200 | 0.34603 |
| 0.9815 | 0.99033 | 31.05 | 200 | 0.89159 |
| 0.9879 | 0.99543 | 85.18 | 200 | 0.75638 |

The $AD$ % values in Table 2 are low. Thus, the maximum value of $AD$ % does not exceed 1.3% for the rigid lattice and 4% for the flexible lattice. These findings confirm the robustness of the 2D model. Although we are comparing a discrete 2D model with a more advanced continuous 3D model, the agreement between the curves in Fig. 4 highlights the significant potential of the discrete 2D sI structure model. It also underscores the effective modeling of the $CH_4$ ($H_2O$) species and the discretization of lateral interactions using Eq. (3).

The study in Figs. 3 and 4 shows that our 2D lattice-gas model aligns well with both experimental results and more advanced 3D simulations. This simple 2D gas model captures key characteristics of the real $CH_4$ clathrate hydrate system and could serve as a valuable tool for studying and describing such systems.

### 3.2. Pressure-temperature phase diagram

In Fig. 5, $\Delta H_{diss}$ [Eqs. (11-13), part (a)] and $\beta f$ [Eq. (10), part (b)] are shown as a function of cavity coverage $\theta_{cavity}$. Four temperatures are presented in the figure: $T = 274, 280, 286$ K (main figure) and $T = 292$ K (inset). As expected, the $\Delta H_{diss}$ curves [Fig. 5(a)] show a similar increase as both large and small cavities are filled at different temperatures. An increase in the chemical potential of the guest species results in greater occupancy of the sI structure, accompanied by an increase in $\Delta H_{diss}$.

Regarding the free energy [Fig. 5(b)], $\beta f$ curves exhibit a minimum around $\theta_{cavity} \approx 1$, indicating the condition where the system is thermodynamically most stable. By projecting this minimum onto Fig. 5(a), the enthalpy value at the more stable condition can be determined (this projection is depicted in the figure using differently colored continuous lines for each temperature $T$). Thus, for each pair ($\theta_{cavity}^{min}$, $\beta f_{min}$) representing the minimum in the free energy curve, there is a corresponding pair of values ($\theta_{cavity}^{min}$, $\Delta H_{diss}^{stable}$) indicating the enthalpy at the most stable condition. The enthalpy values $\Delta H_{diss}^{stable}$ obtained from Fig. 5 remain nearly constant, showing slight variations with temperature: $\Delta H_{diss}^{stable} = 79.3$ kJ/mol (274 K); $\Delta H_{diss}^{stable} = 77.1$ kJ/mol (280 K); $\Delta H_{diss}^{stable} = 73.8$ kJ/mol (286 K); and $\Delta H_{diss}^{stable} = 75.2$ kJ/mol (292 K). Corresponding with the minimum free energy values, the $\Delta H_{diss}^{stable}$ values represent occupancy states where there is approximately one molecule per cavity.

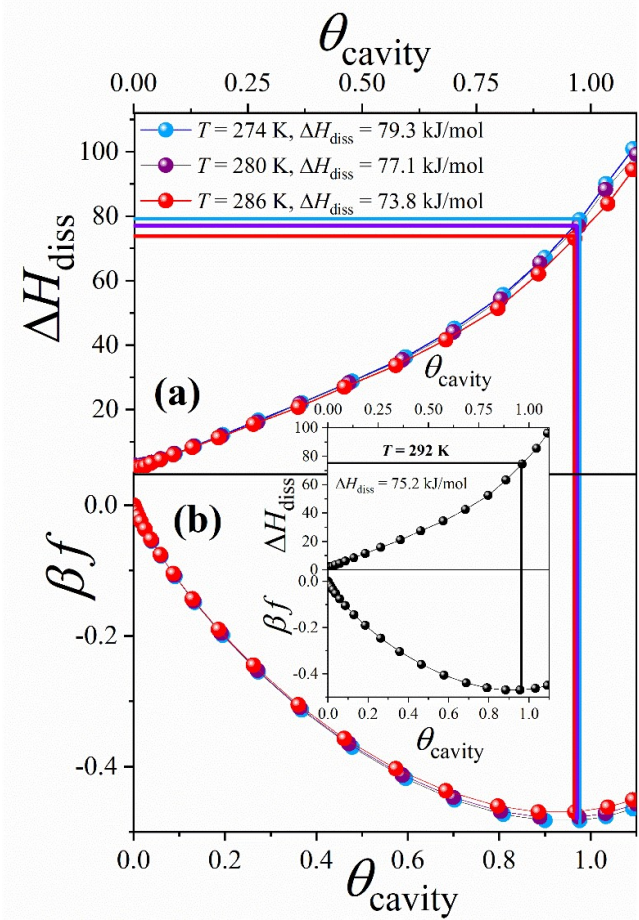

**Fig. 5:** (a) $\Delta H_{diss}$ vs. $\theta_{cavity}$ for three different values of temperature ($T$ = 274, 280, 286 K). The horizontal axis is located at the top of the figure. (b) $\beta f$ as a function of $\theta_{cavity}$ for the same values of $T$ used in part (a). The minimum $\beta f$ values are very close to $\theta_{cavity}$ 1.0. In part (b), the projection of the minimum $\beta f$ values (using continuous lines of different colors for each temperature) with their corresponding $\Delta H_{diss}$ values can be seen in the upper part of figure (a). Inset: Same as the main figure, but for $T$ = 292 K.

The mean value of the dissociation enthalpy $\langle \Delta H_{diss}^{stable} \rangle_{average}$ was calculated by averaging the four determinations obtained in Fig. 5. The resulting value $\langle \Delta H_{diss}^{stable} \rangle_{average}$ = 76.4 kJ/mol differs by 2 per cent from the experimental value in the literature $\Delta H_{exp}$ = 78 kJ/mol [73]. This difference is much smaller than the typical experimental errors associated with these types of measurements, which (1) reinforces the validity of the 2D lattice-gas model, and (2) provides direct evidence that our 2D scheme is a good representation of the real system (CH$_4$ clathrate hydrate with an sI structure). Future efforts will focus on investigating the behavior of our model with other guest species, such as CO$_2$, N$_2$, C$_2$H$_4$, etc., to further enhance the general applicability of the model.

Once $\left\langle \Delta H_{diss}^{stable} \right\rangle_{average}$ was obtained, Eq. (14) was used to calculate the *P-T* phase diagram [73]. The results are shown in Fig. 6: blue spheres correspond to calculations in this work using the 2D lattice-gas model and MC simulations; red open circles correspond to experimental measurements from Refs. [79, 80, 81]; and the black dashed line represents the predictions of a theoretical model derived from the van der Waals-Platteeuw theory. The theoretical approach employs the modified UNIFAC method to calculate activity coefficients in the liquid phase and the PRSV equation of state (EOS) to determine the fugacity of the guest molecules in the vapor phase [26]. It is important to note that the theoretical model presented in Ref. [26] does not include the quadruple point or a small range near it. Nonetheless, we consider the theoretical model in Ref. [26] to be one of the most comprehensive available in the current literature.

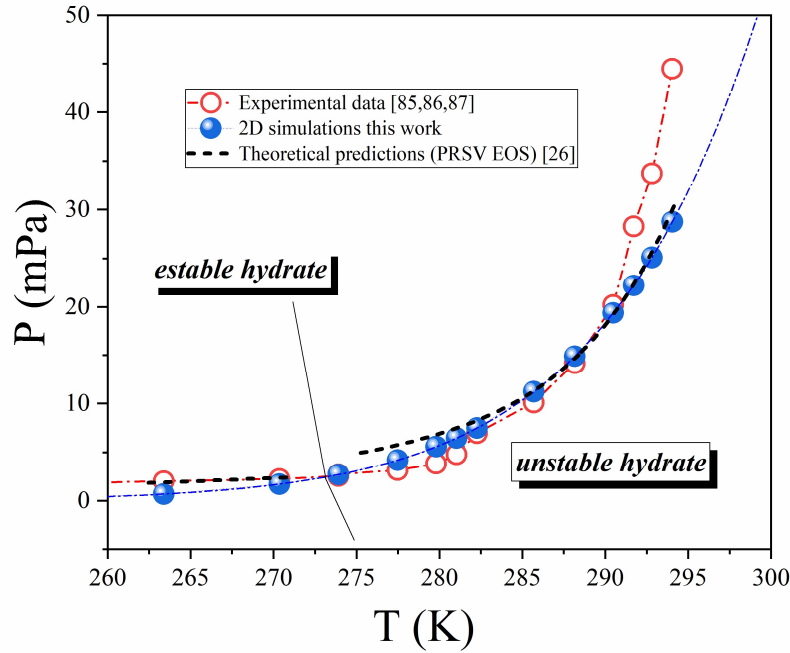

**Fig. 6:** Pressure-temperature phase diagram. Blue spheres correspond to the phase diagram calculated in this work using the 2D lattice-gas model and MC simulations. Red open circles represent experimental measurements [85, 86, 87]. The black dashed line corresponds to the theoretical model using the PRSV equation of state (EOS) [26].

In line with previous results, the 2D phase diagram shows a very good agreement with both the experimental and theoretical phase diagrams shown in Fig. 6. Our model demonstrates solid agreement within the 260-290 K temperature range (corresponding to a pressure range of 2 to 20 MPa) when compared to experimental results. However, for temperatures above 290 K and pressures exceeding 21 MPa, our results deviate from the experimental data points. This behavior is typical of lattice-gas models at higher temperatures, where the lateral couplings between $CH_4$ and $H_2O$ molecules weaken as the system's temperature increases, resulting in reduced sensitivity to the physical behavior of these lateral interactions.

On the other hand, when our simulation results are compared with the theoretical phase diagram, there is a notable agreement across the entire range of $P$ and $T$ in relation to the experimental results. This behavior highlights the validity of the hypothesis that a real hydrate can be represented by an artificial adsorption system consisting of an empty solid structure (adsorbent) in contact with a gas phase of guest molecules (adsorbate).

In summary, the results presented in subsections 3.2 and 3.1 demonstrate the applicability of the proposed 2D lattice-gas model in describing the thermodynamic behavior of real methane clathrate hydrates.

4. **Conclusions**

In the present work, a 2D lattice-gas model combined with Monte Carlo simulations was used to calculate the occupancy isotherms, dissociation enthalpy, and pressure-temperature phase diagram for methane clathrate hydrates with an sI structure. The calculated thermodynamic properties were then compared with experimental data and advanced 3D simulations reported in the current literature. The main findings of this paper are summarized as follows:

- The adsorption isotherms obtained from the 2D lattice-gas model developed in this study were compared with experimental data [75, 76] and 3D simulation results [35, 36, 41]. This comprehensive analysis was conducted at two different temperatures (200 K and 270 K) and included both rigid and flexible sI structures. In the rigid structure case, the aqueous lattice remains fixed during the simulation process, while in the flexible structure case, water molecules are allowed to move

- to neighboring sites. In all cases, the 2D model demonstrated excellent performance. In addition to the study of cavity occupancy, interesting confinement effects were analyzed and discussed in terms of the values of the parameters used in the 2D simulation scheme.
- A simulation method was developed to determine the dissociation enthalpy of the system. The technique involves finding the minimum value of the free energy curve ($\theta_{cavity}^{min} \approx 1.0$) and, from this minimum, calculating the corresponding enthalpy value. The procedure is repeated at various temperatures, and the obtained values are averaged to determine a representative mean value for the dissociation enthalpy of the system $\langle \Delta H_{diss}^{stable} \rangle_{average}$. In this case, $\langle \Delta H_{diss}^{stable} \rangle_{average}$ = 76.4 kJ/mol, which is very close to the experimental result of 78 kJ/mol [73]. The excellent agreement between simulation and experiment demonstrates the reliability of the proposed numerical methodology and its potential use, in conjunction with the Clausius-Clapeyron equation, for calculating *P-T* phase diagrams.
- Taking advantage of the method described in the point above, the pressure-temperature phase diagram of the system was determined using a combination of the Clausius-Clapeyron equation and the calculated average dissociation enthalpy. The results were successfully compared with experimental [79, 80, 81] and theoretical [26] data in the literature.

The results discussed in the previous points illustrate the effectiveness of the proposed 2D lattice-gas model in describing the thermodynamic behavior of real methane clathrate hydrates. Three key characteristics support the strong alignment of our model with the actual hydrate: (i) the symmetry of the sI hydrate structure, which is well-represented by a 2D triangular lattice (see Figs. 1 and 2); (ii) the geometric shape of the methane molecule, which spatially corresponds to an equilateral triangle on the lattice (see Fig. 2); and (iii) the appropriate discretization of the lateral interactions between $CH_4$ and $H_2O$ species. These features effectively capture the fundamental physical essence of real $CH_4$ clathrate hydrate. Future work will concentrate on exploring how our model behaves with other guest species, such as $CO_2$, $N_2$, and $C_2H_4$, etc., to further broaden its general applicability.

**ACKNOWLEDGMENTS**

This work was supported in part by CONICET (Argentina) under project number PIP 11220220100238CO and Universidad Nacional de San Luis (Argentina) under project No. 03-1920. The numerical work was done using the BACO parallel cluster (http://cluster infap.unsl.edu.ar/wordpress/) located at Instituto de Física Aplicada, Universidad Nacional de San Luis - CONICET, San Luis, Argentina. Our work was also supported by ANPCyT through 01-PICT 2022-2022-01-00049 research grant, as well as by SGCyT-UNS. MEP is a member of IFISUR-CONICET. JJ is a fellow researcher at this institution.